\documentclass[aps,prl,twocolumn]{revtex4}
\usepackage{amssymb}
\usepackage{amsmath}
\usepackage{graphicx}
\usepackage{epsfig}

\begin{document}

\title{Quantum information storage and state transfer based on spin systems}
\author{Z. Song$^{1,a}$ and C. P. Sun $^{1,2,a,b}$}
\affiliation{$^{1}$Department of Physics, Nankai University, Tianjin 300071, China}
\affiliation{$^{2}$ Institute of Theoretical Physics, Chinese Academy of Sciences,
Beijing, 100080, China}

\begin{abstract}
The idea of quantum state storage is generalized to describe the coherent
transfer of quantum information through a coherent data bus. In this
universal framework, we comprehensively review our recent systematical
investigations to explore the possibility of implementing the physical
processes of quantum information storage and state transfer by using quantum
spin systems, which may be an isotropic antiferromagnetic spin ladder system
or a ferromagnetic Heisenberg spin chain. Our studies emphasize the physical
mechanisms and the fundamental problems behind the various protocols for the
storage and transfer of quantum information in solid state systems.
\end{abstract}

\maketitle

\section{I. Introduction}

The current development of quantum information science and technology
demands optimal systems serving as long-lived quantum memories, through
which the quantum information carried by a quantum system with short
decoherence time can be coherently transferred \cite{q-inf}. In this sense a
quantum channel or a quantum data bus is needed for perfect transmission of
quantum states. In this article, we will demonstrate that both the quantum
information storage and the quantum state transfer can be uniquely described
in a universal framework.

There exist some schemes \cite{lukin,Flei,sun-prl} concerning about quantum
storage of photon states, while there are also some efforts devoted to the
universal quantum storage for a qubit (a basic two-level system) state,
which is necessary in quantum computation. For example, most recently an
interesting protocol \cite{lukin1,zoller,exp} was presented to reversibly
map the electronic spin state onto the collective spin state of the
surrounding nuclei. Because of the long decoherence time of the nuclear
spins, the information stored in them can be robustly preserved. It was
found that \cite{szs}, only under two homogeneous conditions with low
excitations, such many-nuclei system approximately behaves as a single mode
boson to serve as an efficient quantum memory.

The low excitation condition requires a ground state with all spins
orientated, which can be prepared by applying a magnetic field polarizing
all spins along the same direction. With the concept of spontaneous symmetry
breaking (SSB), one can recognize that a ferromagnetic Heisenberg spin chain
usually has a spontaneous magnetization, which naturally offers such a kind
of ground state. In happen of SSB, the intrinsic interaction between spins
will strongly correlate with the nuclei to form the magnon, a collective
mode of spin wave, even without any external magnetic field. With these
considerations, Wang, Li, Song, and Sun \cite{wlss} explored the possibility
of using a ferromagnetic quantum spin system, instead of the free nuclear
ensemble, to serve as a robust quantum memory. A protocol was present to
implement a quantum storage element for the electronic spin state in a ring
array of interacting nuclei. Under appropriate control of both the electron
and the external magnetic field, an arbitrary quantum state of the
electronic spin qubit, either pure or mixed state, can be coherently stored
in the nuclear spin wave and then read out in reverse process.

On the other hand, designed for a more realistic quantum computing, a
scalable architecture of quantum network should be based on the solid state
system \cite{Div,Bose}. However, the intrinsic feature of solid state based
channels, such as the finiteness of the correlation length \cite%
{White,Dagotto} and the environment induced noise (especially the low
frequency noise) may block this scalability. Fortunately, analytical study
shows that a spin system possessing a commensurate structure of energy
spectrum matched with the corresponding parity can ensure the perfect state
transfer \cite{Matt1,songz1,Matt2}. Based on this fact, an isotropic
antiferromagnetic spin ladder system can be pre-engineered as a novel robust
kind of quantum data bus \cite{songz2}. Because the effective coupling
strength between the two spins connected to a spin ladder is inversely
proportional to the distance of the two spins, the quantum information can
be transferred between the two spins separated by a longer distance. Another
example of the near-perfect transfer of quantum information was given to
illustrate an application of the theorem. The protocol of such near-perfect
quantum state transfer is proposed by using a ferromagnetic Heisenberg chain
with uniform coupling constant, but an external parabolic magnetic field
\cite{songz1}.

The present paper will give a broad overview of the present situation of the
our investigations mentioned above on quantum state storage and quantum
information coherent transfer based on quantum spin systems. We will
understand the physical mechanisms and the fundamental problems behind these
protocols in the view of a unified conception, the generalized quantum
information storage.

\section{II. Generalized Quantum Storage as a Dynamic Process}

For the dynamic process recording and reading quantum information carried by
quantum states, we first describe the idea of generalized quantum storage,
which was also introduced in association with the Berry's phase factor\cite%
{geo}. Let $M$ be a quantum memory possessing a subspace spanned by $%
|M_{n}\rangle ,(n=1,2,...,d$, $\left\langle M_{n}\right. \left\vert
M_{m}\right\rangle =\delta _{nm}),$ which can store the quantum information
of a system $S$ with basis vectors $|S_{n}\rangle ,n=1,2,...,d$. If there
exists a controlled time evolution interpolating between the initial state $%
|S_{n}\rangle $ $\otimes |M\rangle $ and the final state $|S\rangle \otimes
|M_{n}\rangle $ for each index $n$ and arbitrarily given states $|S\rangle $
and $|M\rangle $, we define the usual quantum storage by using a factorized
evolution of time $T_{m}$
\begin{equation}
|\Phi (T_{m})\rangle =U(T_{m})|\Phi (0)\rangle =|S\rangle \otimes
|M_{n}\rangle ,
\end{equation}%
starting from the initial state $|\Phi (0)\rangle =|S_{n}\rangle \otimes
|M\rangle $. The corresponding readout process is an inverse evolution of \
time $T_{f}(>T_{m})$
\begin{equation}
|\Phi (T_{f})\rangle =U(T_{f})|\Phi (0)\rangle =|S_{n}\rangle \otimes
|M\rangle.
\end{equation}
In this sense, writing an arbitrary state $|S(0)\rangle
=\sum_{n}c_{n}\,|S_{n}\rangle $ of $S$ into $M$ with the initial state $%
|M\rangle $ of quantum memory can be realized as a controlled evolution from
time $t=0$ to $t=T_{m}$
\begin{equation}
\sum_{n}c_{n}|S_{n}\rangle \otimes |M\rangle \rightarrow |S\rangle \otimes
\sum_{n}c_{n}|M_{n}\rangle.
\end{equation}%
The readout process from $M$ is another controlled evolution from time $%
t=T_{m}$ to $t=T_{f}$
\begin{equation}
|S\rangle \otimes \sum_{n}c_{n}|M_{n}\rangle \rightarrow
\sum_{n}c_{n}|S_{n}\rangle \otimes |M\rangle.
\end{equation}
Obviously, the combination of these two processes forms a cyclic evolution
that a state totally returns to the initial one.

However, in the view of the decoding approach, one need not the
\textquotedblleft totally returning" to revival the information of initial
state and a difference is allowed by $n-independent$ unitary transformation $%
W=W_{S}$ $\otimes 1$, namely,
\begin{equation}
|S\rangle \otimes W_{M}\sum_{n}c_{n}|M_{n}\rangle \rightarrow
(W_{S}\sum_{n}c_{n}|S_{n}\rangle )\otimes |M\rangle .
\end{equation}%
This is a  quantum dynamic process for  recording and reading, which defines
a quantum storage. Because the factor $W_{S}$ is known to be independent of
the initially state, it can be easily decoded from $W_{S}\sum_{n}c_{n}|S_{n}%
\rangle $ by the inverse transformation of $W_{S}$.
\begin{figure}[tbp]
\includegraphics[bb=90 280 520 580, width=7 cm, clip]{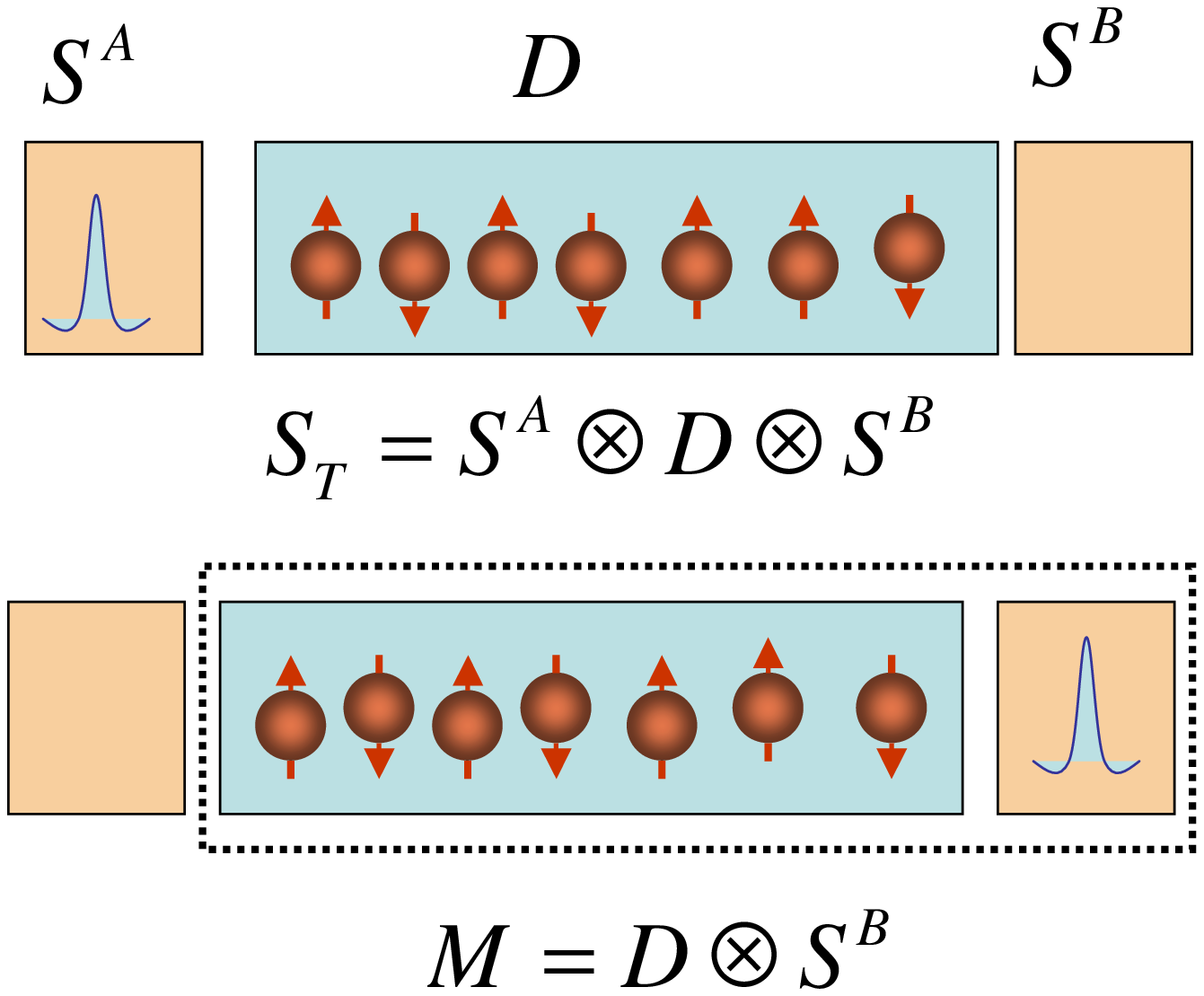}
\caption{Demonstration of quantum state transfer as a process of generalized
quantum information storage by grouping the data bus $D$ and the target
subsystem $S^{B}$as a generalized quantum memory.}
\end{figure}
We notice that the quantum storage usually relates to two quantum
subsystems.

\bigskip

We will show as follows that the quantum state transfer can be understood as
a generalized quantum storage with three subsystems, the input one with the
Hilbert space $S^{A}$, the data bus with $D$ and output one with $S^{B}$. As
illustrated in Fig. 1, the two subsystems $S^{A}$ and $S^{B}$ located at two
distant locations $A$ and $B$ respectively. Then the Hilbert space of the
total system can be written as
\begin{equation}
S_{T}=S^{A}\otimes D\otimes S^{B}\equiv S^{A}\otimes M,
\end{equation}%
where $M=D\otimes S^{B}$ can be regarded as the generalized quantum memory
with the memory space spanned by $|M_{n}\rangle =\left\vert D\right\rangle
\otimes U_{B}\left\vert S_{n}^{B}\right\rangle $. Here $\left\vert
D\right\rangle $ is a robust state of the data bus and $U_{B}$ represents
some local unitary transformations with respect to $B$, which are
independent of the initial state. With this notation, the quantum state
transfer indeed can be regarded as a generalized QDR.

In fact, if one input a state of $\left\vert S^{A}\right\rangle
=\sum\nolimits_{n}c_{n}\left\vert S_{n}^{A}\right\rangle $ localized at $A$
at $t=0$, the initial state of whole system can be written as
\begin{equation}
\left\vert \psi (0)\right\rangle =\sum\nolimits_{n}c_{n}\left\vert
S_{n}^{A}\right\rangle \otimes |M\rangle
\end{equation}%
where $|M\rangle =\left\vert D\right\rangle \otimes \left\vert
S^{B}\right\rangle $. The quantum state transfer can be usually described as
a factorized time evolution at time $t=T_{f}$
\begin{eqnarray}
\left\vert \psi (T_{f})\right\rangle  &=&\left\vert S\right\rangle \otimes
\left\vert D\right\rangle \otimes \sum\nolimits_{n}c_{n}U_{B}\left\vert
S_{n}^{B}\right\rangle   \notag \\
&=&|S\rangle \otimes \sum\nolimits_{n}c_{n}|M_{n}\rangle
\end{eqnarray}%
with $|M_{n}\rangle =\left\vert D\right\rangle \otimes U_{B}\left\vert
S_{n}^{B}\right\rangle $. The above equations just demonstrate that the
quantum state transfer is essentially a generalized quantum memory with $%
W_{M}=(1\otimes U_{B})$. In this sense the revisable quantum state transfer
can be regarded as a general readout process.

Now we would like to remark on the differences between generalized quantum
state storage and other two types of quantum processes, quantum
teleportation and quantum copy. In fact, quantum teleportation is
theoretically perfect, yielding an output state which revival the input with
a fidelity $F=1$. Actually one of necessary procedure in teleportation is to
measure the Bell state at location $A$, which will induce the collapse of
wavepacket. On the other way around, the quantum state storage process is
always on time evolution without any measurement. As for quantum copy the
initial state remains unchanged during its copy can be generated in a
dynamic process.

\section{III. Quantum state transfer in spin systems}

A robust quantum information processing based on solid state system is
usually implemented in a working spaces panned by the lowest states, which
are well separated from other dense spectrum of high excitations. In this
sense the energy gap of the solid state system is an important factor we
should take into account. The decoherence induced by the environmental noise
can also destroy the robustness of quantum information processing, such as
the low frequency (e.g, $1/f$ ) noise dominating in the solid state devices.
People believe that the gap of the data bus can suppress the stay of
transferred state in the middle way in order to enhance the fidelity, but
the large gap may result in a shorter correlation length. The relationship
between correlation length and the energy gap is usually established in the
system with translational symmetry. So we need to consider some
modulated-coupling systems or artificially engineered irregular quantum spin
systems where the strong correlation between two distant site can be
realized.

\subsection{1. Theorem for the perfect quantum state transfer}

Quantum mechanics shows that perfect state transfer is possible. To sketch
our central idea, let us first consider a single particle system with the
usual spatial refection symmetry (SRS) in the Hamiltonian $H$. Let $P$ be
the spatial refection operator. The SRS is implied by $[H,P]=0$. Now we
prove that at time $\pi /E_{0}$ any state $\psi (\mathbf{r})$ can evolve
into the reflected state $\pm \psi (-\mathbf{r})$ if the eigenvalues $%
\varepsilon _{n}$ match the parities $p_{n}$ in the following way
\begin{equation}
\varepsilon _{n}=N_{n}E_{0},p_{n}=\pm (-1)^{N_{n}}  \label{spmc}
\end{equation}%
for arbitrary positive integer $N_{n}$ and
\begin{equation}
H\phi _{n}(\mathbf{r})=\varepsilon _{n}\phi _{n}(\mathbf{r}),P\phi _{n}(%
\mathbf{r})=p_{n}\phi _{n}(\mathbf{r}).
\end{equation}%
Here, $\phi _{n}(\mathbf{r})$ is the common eigen wave function of $H$ and $P
$, $\mathbf{r}$ is the position of the particle. We call Eq. (\ref{spmc})
the spectrum-parity matching condition (SPMC). The proof of the above
rigorous conclusion is a simple, but heuristic exercise in basic quantum
mechanics. In fact, for the spatial refection operator, $P\psi (\mathbf{r}%
)=\pm \psi (-\mathbf{r})$. For an arbitrarily given state at $t=0,$ $\psi (%
\mathbf{r},t)\left\vert _{t=0}\right. =\psi (\mathbf{r})$, it evolves to
\begin{equation}
\psi (\mathbf{r},t)=e^{-iHt}\psi (\mathbf{r})=\sum_{n}C_{n}e^{-iN_{n}E_{0}t}%
\phi _{n}(\mathbf{r})
\end{equation}%
at time $t$, where $C_{n}=\left\langle \phi _{n}\right\vert \psi \rangle $.
Then at time $t=\pi /E_{0}$, we have
\begin{equation}
\psi (\mathbf{r},\frac{\pi }{E_{0}})=\sum_{n}C_{n}(-1)^{N_{n}}\phi _{n}(%
\mathbf{r})=\pm P\psi (\mathbf{r})
\end{equation}%
that is $\ \psi (\mathbf{r},\pi /E_{0})=\pm \psi (-\mathbf{r}).$This is just
the central result \cite{MIS} discovered for quantum spin system that the
evolution operator becomes a parity operators $\pm P$ at some instant $%
t=(2n+1)\pi /E_{0}$, that is $\exp [-iH\pi /E_{0}]=\pm P.$From the above
arguments we have a consequence that if the eigenvalues $\varepsilon
_{n}=N_{n}E_{0}$ of a 1-D Hamiltonian $H$ with spatial refection symmetry
are odd-number spaced, i.e. $N_{n}-N_{n-1}$ are always odd, any initial
state $\psi (x)$ can evolve into $\pm \psi (-x)$ at time $t=\pi /E_{0}$. In
fact, for such 1-D systems, the discrete states alternate between even and
odd parities. Consider the eigenvalues $\varepsilon _{n}=N_{n}E_{0}$ with
odd-number spaced. The next nearest level must be even-number spaced, then
the SPMC is satisfied. Obviously, the 1-D SPMC is more realizable for the
construction of the model Hamiltonian to perform perfect state transfer.

Now, we can directly generalize the above analysis to many particle systems.
\ For the quantum spin chain, one can identify the above SRS as the MIS with
respect to the center of the quantum spin chain. As the discussion in Ref.
\cite{MIS}, we write MIS operation
\begin{equation}
P\Psi (s_{1,}s_{2,}...,s_{N-1,}s_{N})=\Psi (s_{N},s_{N-1,}...,s_{2},s_{1,})
\end{equation}%
for the wave function $\Psi (s_{1,}s_{2,}...,s_{N-1,}s_{N})$ of spin chain.
Here, $s_{n}=0,1$ denotes the spin values of the n-th qubit.

\subsection{2. Perfect state transfer in modulated coupling system}

Based on the above analysis, in principle, perfect quantum state transfer is
possible in the framework of quantum mechanics. According to SPMC, many spin
systems can be pre-engineered for perfect quantum states transfer. For
instance, two-site spin-$\frac{1}{2}$ Heisenberg system is the simplest
example which meets the SPMC. Recently, M. Christandl et al \cite%
{Matt1,Matt2} proposed a $N$-site $XY$ chain with an elaborately designed
modulated coupling constants between two nearest neighbor sites, which
ensures a perfect state transfer. It is easy to find that this model
corresponds the SPMC for the simplest case $N_{n}=n$. A natural extension of
the application of the theorem leads to discover other models with $%
N_{n}\neq n$. Following this idea, a new class of different models whose
spectrum structures obey the SPMC exactly were proposed for perfect state
transfer. Consider an $N$-site spin$-\frac{1}{2}$ $XY$ chain with the
Hamiltonian
\begin{equation}
H=2\sum_{i=1}^{N-1}J_{i}[S_{i}^{x}S_{i+1}^{x}+S_{i}^{y}S_{i+1}^{y}]
\end{equation}%
where $S_{i}^{x},S_{i}^{y}$ and $S_{i}^{z}$ are Pauli matrices for the $i-$%
th site, $J_{i}$ is the coupling strength for nearest neighbor interaction.
For the open boundary condition, this model is equivalent to the spin-less
fermion model. The equivalent Hamiltonian can be written as
\begin{equation}
H=\sum\limits_{i=1}^{N-1}J_{i}^{[k]}a_{i}^{\dag }a_{i+1}+h.c,
\end{equation}%
where $a_{i}^{\dag },a_{i}$ are the fermion operators. This describes a
simple hopping process in the lattice. According to the SPMC, we can present
different \ models (labelled by different positive integer $k\in
\{0,1,2,...\}$) through pre-engineering of the coupling strength as $%
J_{i}=J_{i}^{[k]}=\sqrt{i\left( N-i\right) }$for even $i$ \ and $%
J_{i}=J_{i}^{[k]}=\sqrt{\left( i+2k\right) \left( N-i+2k\right) }$for odd $i$%
. By a straightforward calculation, one can find the k-dependent spectrum$%
\varepsilon _{n}=-N+2(n-k)-1$ for $n=1,2,...,N/2,$and $\varepsilon _{n}=-N+$
$2(n+k)-1$ for $n=N/2+1,...,N$. The corresponding k-dependent eigenstates
are
\begin{equation}
\left\vert \phi _{n}\right\rangle =\sum_{i=1}^{N}c_{ni}\left\vert
i\right\rangle =\sum_{i=1}^{N}c_{ni}a_{i}^{\dag }\left\vert 0\right\rangle
\end{equation}%
where the coefficients $c_{ni}$ can be explicitly determined by the
recurrence relation presented in Ref. \cite{songz2}.

It is obvious that the model proposed in Ref. \cite{Matt1} is just the
special case of our general model in $k=0$. For arbitrary $k$, one can
easily check that it meets the our SPMC by a straightforward calculation.
Thus we can conclude that these spin systems with a set of pre-engineered
couplings $J_{i}^{[k]}$ can serve as the perfect quantum channels that allow
the qubit information transfer.

\subsection{3. Near-perfect state transfer}

In real many-body systems, the dimension of Hilbert space increase with the
size $N$ exponentially. For example, $N$-site spin-$\frac{1}{2}$ system, the
dimension is $D=2^{N}$, and the symmetry of the Hamiltonian can not help so
much. So it is almost impossible to obtain a model to be exactly engineered.
In the above arguments we just show the possibility to implement the perfect
state transfer of any quantum state over arbitrary long distances in a
quantum spin chain. It sheds light into the investigation of near-perfect
quantum state transfer. There is a naive way that one select some special
states to be transported, which is a coherent superposition of commensurate
part of the whole set of eigenstates. For example, we consider a truncated
Gaussian wavepacket for an anharmonic oscillator with lower eigenstates to
be harmonic. It is obvious that such system allows some special states to
transfer with high fidelity. We can implement such approximate harmonic
system in a natural spin chain without the pre-engineering of couplings, but
the present of a modulated external field. Another way to realize near
perfect state transfer is\ to achieve the entangled states and fast quantum
states transfer of two spin qubits by connecting two spins to a medium which
possesses a spin gap. A perturbation method, the Fr\H{o}hlich
transformation, shows that the interaction between the two spins can be
mapped to the Heisenberg type coupling.

\subsubsection{3.1 Spin ladder}

We sketch our idea with the model illustrated in Fig. 2. The whole quantum
system we consider here consists of two qubits (A and B) and a $2\times N$%
-site two-leg spin ladder. In practice, this system can be realized by the
engineered array of quantum dots \cite{QD array}. The total Hamiltonian $%
H=H_{M}+H_{q}$
\begin{figure}[h]
\includegraphics[bb=40 240 550 660, width=7 cm, clip]{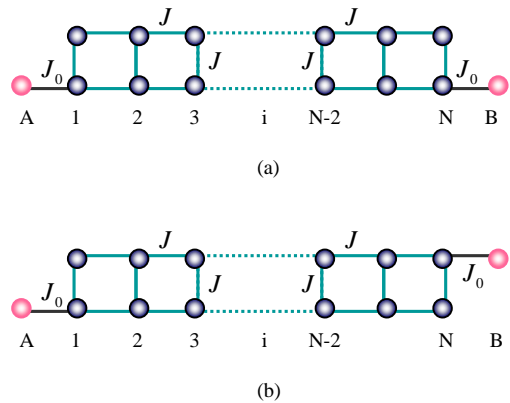}
\caption{Two qubits $A$ and $B$ connect to a $2\times N$-site spin ladder.
The ground state of $H$ with a-type connection (a) is singlet (triplet) when
$N$ is even (odd), while for b-type connection (b), one should have opposite
result.}
\end{figure}
contains two parts, the medium Hamiltonian
\begin{equation}
H_{M}=J\sum_{\left\langle ij\right\rangle \perp }\mathbf{S}_{i}\cdot \mathbf{%
S}_{j}+J\sum_{\left\langle ij\right\rangle \parallel }\mathbf{S}_{i}\cdot
\mathbf{S}_{j}  \label{2}
\end{equation}%
describing the spin-1/2 Heisenberg spin ladder consisting of two coupled
chains and the coupling Hamiltonian
\begin{equation}
H_{q}=J_{0}\mathbf{S}_{A}\cdot \mathbf{S}_{L}+J_{0}\mathbf{S}_{B}\cdot
\mathbf{S}_{R}  \label{3}
\end{equation}%
describing the connections between qubits $A$, $B$ and the ladder. In the
term $H_{M},$ $i$ denotes a lattice site on which one electron sits, $%
\left\langle ij\right\rangle $ $\perp $ denotes nearest neighbor sites on
the same rung, $\left\langle ij\right\rangle $ $\parallel $ denotes nearest
neighbors on either leg of the ladder. In term $H_{q}$, $L$ and $R$ denote
the sites connecting to the qubits $A$ and $B$ at the ends of the ladder.
There are two types of the connection between $\mathbf{S}_{A}(\mathbf{S}_{B})
$ and the ladder, which are illustrated in Fig. 2. According to the Lieb's
theorem \cite{Lieb}, the spin of the ground state of $H$ with the connection
of type a is zero (one) when $N$ is even (odd), while for the connection of
type b, one should have an opposite result. For the two-leg spin ladder $%
H_{M},$ analytical analysis and numerical results have shown that the ground
state and the first excited state of the spin ladder have spin $0$ and $1$
respectively \cite{Dagotto,White}. It is also shown that there exists a
finite spin gap $\triangle =E_{1}^{M}-E_{g}^{M}\sim J/2.$between the ground
state and the first excited state (see the Fig. 3). This fact has been
verified by experiments \cite{Dagotto} and is very crucial for our present
investigation.
\begin{figure}[h]
\includegraphics[bb=45 320 540 650, width=7 cm, clip]{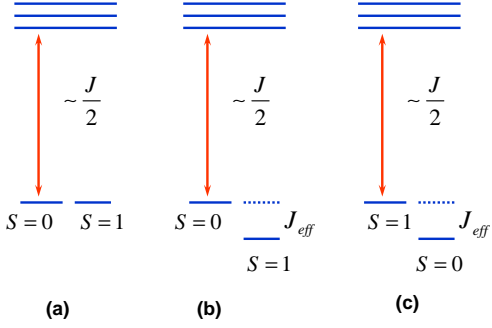}
\caption{Schematic illustration of the energy levels of the system. (a) When
the connections between two qubits and the medium switch off ($J_{0}=0$) the
ground states are degenerate. (b), (c) When $J_{0}$ switches on, the ground
state(s) and the first excited state(s) are either singlet or triplet. This
is approximately equivalent to that of two coupled spins. }
\end{figure}
Thus, it can be concluded that the medium can be robustly frozen to its
ground state to induce the effective Hamiltonian $H_{eff}=J_{eff}\mathbf{S}%
_{A}\cdot \mathbf{S}_{B}$ between the two end qubits. With the effective
coupling constant $J_{eff}$ to be calculated in the following, this
Hamiltonian depicts the direct exchange coupling between two separated
qubits. As the famous Bell states, $H_{eff}$ has singlets and triplets
eigenstates $\left\vert j,m\right\rangle _{AB}$: $\left\vert
0,0\right\rangle =\frac{1}{\sqrt{2}}\left( \left\vert \uparrow \right\rangle
_{A}\left\vert \downarrow \right\rangle _{B}-\left\vert \downarrow
\right\rangle _{A}\left\vert \uparrow \right\rangle _{B}\right) $ and $%
\left\vert 1,1\right\rangle =\left\vert \uparrow \right\rangle
_{A}\left\vert \uparrow \right\rangle _{B}$, $\left\vert 1,-1\right\rangle
=\left\vert \downarrow \right\rangle _{A}\left\vert \downarrow \right\rangle
_{B}$, $\left\vert 1,0\right\rangle =\frac{1}{\sqrt{2}}\left( \left\vert
\uparrow \right\rangle _{A}\left\vert \downarrow \right\rangle
_{B}+\left\vert \downarrow \right\rangle _{A}\left\vert \uparrow
\right\rangle _{B}\right) $, which can be used as a channel to share
entanglement for a perfect quantum communication in a longer distance.

The above central conclusion can be proved with both analytical and
numerical methods as follows. To deduce the above effective Hamiltonian we
use $|\psi _{g}\rangle _{M}$ ($|\psi _{\alpha }\rangle _{M}$) and $E_{g}$ ($%
E_{\alpha }$) to denote ground (excited) states of $H_{\mathrm{M}}$ and the
corresponding eigen-values. The zero order eigenstates $\left\vert
m\right\rangle $ can then be written as in a joint way
\begin{eqnarray}
\left\vert j,m\right\rangle _{g} &=&\left\vert j,m\right\rangle _{AB}\otimes
|\psi _{g}\rangle _{M},  \notag \\
\left\vert \psi _{\alpha }^{jm}(s^{z})\right\rangle  &=&\left\vert
j,m\right\rangle _{AB}\otimes |\psi _{\alpha }\rangle _{M}
\end{eqnarray}%
Here, we have considered that $z$-component $%
S^{z}=S_{M}^{z}+S_{A}^{z}+S_{B}^{z}$ of total spin is conserved with respect
to the connection Hamiltonian $H_{q}$. Since $S_{M}^{z}$ and $S_{M}^{2}$
commute with $H_{M}$, we can label $|\psi _{g}\rangle _{M}$ as $|\psi
_{g}(s_{M},s_{M}^{z},)\rangle _{M}$ and then $s^{z}=m+s_{M}^{z}$ can
characterize the non-coupling spin state $\left\vert \psi _{\alpha
}^{jm}(s^{z})\right\rangle $.

When the connections between the two qubits and the medium are switched off,
i.e., $J_{0}=0,$ the degenerate ground states of $H$ are just $\left\vert
j,m\right\rangle _{g}$ with the degenerate energy $E_{g}$ and spin $0,1$
respectively, which is illustrated in Fig. 3(a). When the connections
between the two qubits and the medium are switched on, the degenerate states
with spin $0,1$ \cite{Song} should split as illustrated in Fig. 3(b) and
(c). In the case with $J_{0}\ll J$ at lower temperature $kT<J/2$, the medium
can be frozen to its ground state and then we have the effective Hamiltonian
\begin{eqnarray}
H_{\mathrm{eff}} &\cong &\sum_{j^{\prime },m^{\prime },j,m,s^{z}}\frac{%
|_{g}\left\langle j,m\right\vert H_{\mathrm{q}}\left\vert \psi _{\alpha
}^{j^{\prime }m^{\prime }}(s^{z})\right\rangle |^{2}}{E_{g}-E_{\mathrm{%
\alpha }}}\left\vert j,m\right\rangle _{gg}\left\langle j,m\right\vert
\notag \\
&=&J_{eff}.\mathit{Diag}.(\frac{1}{4},\frac{1}{4},\frac{1}{4},-\frac{3}{4}%
)+\varepsilon
\end{eqnarray}
where%
\begin{eqnarray}
J_{eff} &=&\sum\limits_{\alpha }\frac{J_{0}^{2}[L(\alpha )R^{\ast }(\alpha
)+R(\alpha )L^{\ast }(\alpha )]}{E_{g}-E_{\alpha }},  \label{10} \\
\varepsilon &=&\sum\limits_{\alpha }\frac{3J_{0}^{2}\left[ \left\vert
L(\alpha )\right\vert ^{2}+\left\vert R(\alpha )\right\vert ^{2}\right] }{%
4\left( E_{g}-E_{\alpha }\right) }.  \notag
\end{eqnarray}%
This just proves the above effective Heisenberg Hamiltonian
$H_{eff}$. Here, the matrix elements of interaction $K(\alpha
)=_{M}\left\langle \psi _{g}\right\vert S_{K}^{z}\left\vert \psi
_{\alpha }\left( 1,0\right) \right\rangle _{M}$ ($K=S,L)$ can be
calculated only for the variables of data bus medium. We also
remark that, because $S^{z}$ and $S^{2}$ are conserved for
$H_{q},$ off-diagonal elements in the above effective Hamiltonian
vanish.

In temporal summary, we have shown that at lower temperature $kT<J/2$, $H$
can be mapped to the effective Hamiltonian $H_{eff}$, which seemingly
depicts the direct exchange coupling between two separated qubits. Notice
that the coupling strength has the form $J_{eff}\sim g(L)J_{0}^{2}/J$, where
$g(L)$ is a function of $L=N+1$, the distance between the two qubits we
concerned. Here we take the $N=2$ case as an example. According to Eq. (\ref%
{10}) one can get $J_{eff}=-(1/4)J_{0}^{2}/J$ and $(1/3)J_{0}^{2}/J$ when $A$
and $B$ connect the plaquette diagonally and adjacently, respectively. This
result is in agreement with the theorem \cite{Lieb} about the ground state
and the numerical result when $J_{0}\ll J$. In general cases, the behavior $%
g(L)$ vs $L$ is very crucial for quantum information since $L/\left\vert
J_{eff}\right\vert $ determines the characteristic time of quantum state
transfer between the two qubits $A$ and $B$. In order to investigate the
profile of $g(L)$, a numerical calculation is performed for the systems $%
L=4,5,6,7,8,$ and $10$, with $J=10,20,40,$ and $J_{0}=1$. The spin gap
between the ground state(s) and first excite state(s) are calculated, which
corresponds to the magnitude of $J_{eff}$. The numerical result is plotted
in Fig. 4, which indicates that $J_{eff}\sim 1/(LJ)$. It implies that the
characteristic time of quantum state transfer linearly depends on the
distance and then guarantees the possibility to realize the entanglement of
two separated qubits in practice.

In order to verify the validity of the effective Hamiltonian $H_{eff}$, we
need to compare the eigenstates of $H_{eff}$ with those reduced states from
the eigenstates of the whole system. In general the eigenstates of $H$ can
be written formally as
\begin{equation}
\left\vert \psi \right\rangle =\sum_{jm}c_{jm}\left\vert j,m\right\rangle
_{AB}\otimes |\beta _{jm}\rangle _{M}
\end{equation}%
where \{$|\beta _{jm}\rangle _{M}$ \} is a set of vectors of the data bus,
which is not necessarily orthogonal. Then we have the condition $%
\sum_{jm}|c_{jm}|_{M}^{2}\langle \beta _{jm}|\beta _{jm}\rangle _{M}=1$ for
normalization of $\left\vert \psi \right\rangle $. In this sense the
practical description of the A-B subsystem of two quits can only be given by
the reduced density matrix
\begin{figure}[h]
\includegraphics[bb=30 300 510 760, width=7 cm, clip]{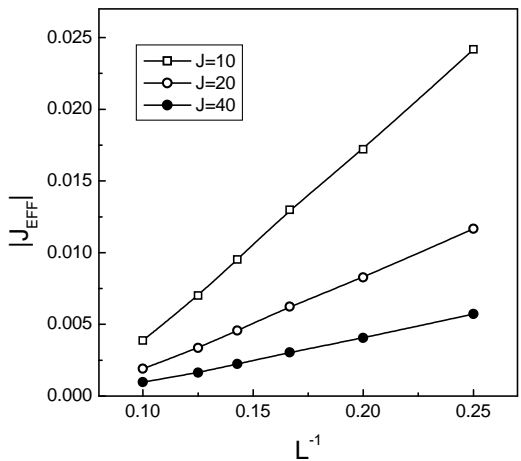}
\caption{The spin gaps obtained by numerical method for the systems $%
L=4,5,6,7,8,$ and $10$, with $J=10,20,40,$ and $J_{0}=1$ are potted, which
is corresponding to the magnitude of $J_{eff}$. It indicates that $%
J_{eff}\sim 1/(LJ)$.}
\end{figure}
\begin{eqnarray}
\rho _{AB} &=&Tr_{M}(\left\vert \psi \right\rangle \left\langle \psi
\right\vert )=\sum_{jm}|c_{jm}|^{2}\left\vert j,m\right\rangle _{AB}\langle
j,m| \\
&&+\sum_{j^{\prime }m^{\prime }\neq jm}c_{j^{\prime }m^{\prime }}^{\ast
}c_{jmM}\langle \beta _{j^{\prime }m^{\prime }}|\beta _{jm}\rangle
_{M}\left\vert j,m\right\rangle _{AB}\langle j^{\prime },m^{\prime }|  \notag
\end{eqnarray}%
where $Tr_{M}$ means the trace-over of the variables of the medium. By a
straightforward calculation we have
\begin{eqnarray}
\left\vert c_{11}\right\vert ^{2} &=&\left\vert c_{1-1}\right\vert
^{2}=\left\langle \psi \right\vert \left( \frac{1}{4}+S_{A}^{z}\cdot
S_{B}^{z}\right) \left\vert \psi \right\rangle ,  \notag \\
\left\vert c_{00}\right\vert ^{2} &=&\left\langle \psi \right\vert \left(
\frac{1}{4}-\mathbf{S}_{A}\cdot \mathbf{S}_{B}\right) \left\vert \psi
\right\rangle , \\
\left\vert c_{10}\right\vert ^{2} &=&1-2\left\vert c_{11}\right\vert
^{2}-\left\vert c_{00}\right\vert ^{2}.  \notag
\end{eqnarray}
Now we need a criteria to judge how close the practical reduced eigenstate
is to the pure state for the effective two sites coupling $H_{eff}$. As we
noticed, it has the singlet and triplet eigenstates $\left\vert
j,m\right\rangle _{AB}$ in the subspace spanned by $\left\vert
0,0\right\rangle _{AB}$ with $S^{z}=S_{A}^{z}+S_{B}^{z}=0$, we have $%
\left\vert c_{11}\right\vert ^{2}=\left\vert c_{10}\right\vert
^{2}=\left\vert c_{1-1}\right\vert ^{2}=0,$ $\left\vert c_{00}\right\vert
^{2}=1;$ for triplet eigenstate $\left\vert 1,0\right\rangle _{AB}$, we have
$\left\vert c_{11}\right\vert ^{2}=\left\vert c_{1-1}\right\vert
^{2}=\left\vert c_{00}\right\vert ^{2}=0,$ $\left\vert c_{10}\right\vert
^{2}=1$. With the practical Hamiltonian $H,$ the values of $\left\vert
c_{jm}\right\vert ^{2}, i=1,2,3,4$ are numerically calculated for the ground
state $\left\vert \psi _{g}\right\rangle $ and first excited state $%
\left\vert \psi _{1}\right\rangle $ of finite system systems $L=4,5,6,7,8$
and $10$ with $J=10,20,$ and $40,$ $(J_{0}=1)$ in $S^{z}=0$ subspace, which
are listed in the Table 1(a,b,c) of Ref. \cite{songz1}. It shows that, at
lower temperature, the realistic interaction leads to the results about $%
\left\vert c_{jm}\right\vert ^{2}$, which are very close to that described
by $H_{eff},$ even if $J$ is not so large in comparison with $J_{0}$.

We address that the above tables reflect all the facts distinguishing the
difference between the results about the entanglement of two end qubit
generated by $H_{eff}$ and $H.$ Though we have ignored the off-diagonal
terms in the reduced density matrix, the calculation of the fidelity $%
F(|j,m\rangle )\equiv ._{M}\langle j,m|\rho _{AB}|j,m\rangle
_{M}=|c_{jm}|^{2}$ further confirms our observation, that the effective
Heisenberg type interaction of two end qubits can approximates the realistic
Hamiltonian very well. Then the quantum information can be transferred
between the two ends of the $2\times N$-site two-leg spin ladder, that can
be regarded as the channel to share entanglement with separated Alice and
Bob. Physically, this is just due to a large spin gap existing in such a
perfect medium, whose ground state can induce a maximal entanglement of the
two end qubits. We also pointed out that our analysis is applicable for
other types of medium systems as data buses, which possess a finite spin
gap. Since $L/\left\vert J_{eff}\right\vert $ determines the characteristic
time of quantum state transfer between the two qubits, the dependence of $%
J_{eff}$ upon $L$ becomes important and relies on the appropriate choice of
the medium.

In conclusion, we have presented and studied in detail a protocol to quantum
state transfer. Numerical results show that the isotropic antiferromagnetic
spin ladder system is a perfect medium through which the interaction between
two separated spins is very close to the Heisenberg type coupling with a
coupling constant inversely proportional to the distance even if the spin
gap is not so large comparing to the couplings between the input and output
spins with the medium.

\subsubsection{3.2 Spin chain in modulated external magnetic field}

Let us consider the Hamiltonian of $(2N+1)$-site spin-$\frac{1}{2}$
ferromagnetic Heisenberg chain
\begin{equation}
H=-J\sum_{i=1}^{2N}\mathbf{S}_{i}\cdot \mathbf{S}_{i+1}+%
\sum_{i=1}^{2N+1}B(i)S_{i}^{z}
\end{equation}%
with the uniform coupling strength $-J<0,$ but in the parabolic magnetic
field
\begin{equation}
B(i)=2B_{0}(i-N-1)^{2}
\end{equation}%
where $B_{0}$ is a constant. In single-excitation invariant subspace with
the fixed z-component of total spin $S^{z}=N-1/2$, this model is equivalent
to the spin-less fermion hopping model with the Hamiltonian
\begin{equation}
H=-\frac{J}{2}\sum\limits_{i=1}^{2N}(a_{i}^{\dag }a_{i+1}+h.c)+\frac{1}{2}%
\sum_{i=1}^{2N+1}B(i)a_{i}^{\dag }a_{i}
\end{equation}%
where we have neglected a constant in the Hamiltonian for simplicity. For
the single-particle case with the set basis$\{\overset{\qquad \ \ \ \ \ \ \
n^{\prime }th}{\left\vert n\right\rangle =\left\vert
0,0,...,1,0...\right\rangle }|n=1,2,..\}$
\begin{figure}[h]
\includegraphics[bb=75 350 525 770, width=7 cm, clip]{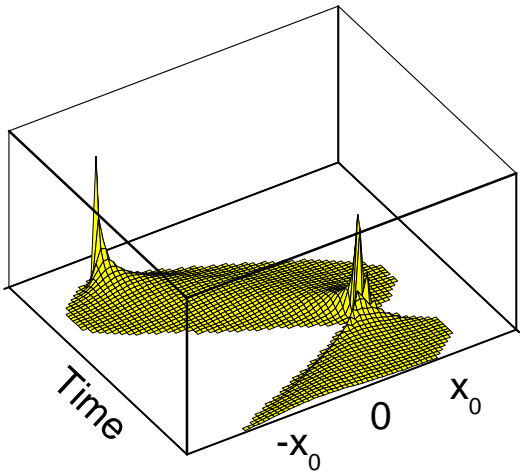}
\caption{Schematic illustration of the time evolution of a Gaussian
wavepacket. It shows that the near-perfect state transfer over a long
distance is possible in the quasi-harmonic system.}
\end{figure}
which is just the same as that of the Hamiltonian of Josephson junction in
the Cooper-pair number basis Ref. \cite{Shn} for $E_{J}=J,E_{c}=2B_{0}$.
Analytical analysis and numerical results have shown that the lower energy
spectrum is indeed quasi-harmonic in the case $E_{J}\gg E_{c}$ \cite{Shi}.
Although the eigenstates of the Hamiltonian (36) does not satisfy the SPMC
precisely, especially for high energy range, there must exist some Gaussian
wavepacket states expanded by the lower eigenstates. Such kind of state can
be transferred with high fidelity.

We consider a Gaussian wavepacket at $t=0$, $x=N_{A}$ as the initial state
\begin{equation}
\left\vert \psi (N_{A},0)\right\rangle =C\sum_{i=1}^{2N+1}e^{-\frac{1}{2}%
\alpha ^{2}(i-N_{A}-1)^{2}}\left\vert i\right\rangle
\end{equation}%
where $\left\vert i\right\rangle $ denotes the state with $2N$ spins in down
state and only the $i$th spin in up state, $C$ is the normalization factor.
The coefficient $\alpha ^{2}=4\ln 2/\Delta ^{2}$ is determined by the width
of the Gaussian wavepacket $\Delta $. The state $\left\vert \psi
(0)\right\rangle $ evolves to $\left\vert \psi (t)\right\rangle
=e^{-iHt}\left\vert \psi (N_{A},0)\right\rangle $ at time $t$ and the
fidelity for the state $\left\vert \psi (0)\right\rangle $ transferring to
the position $N_{B}$ is defined as
\begin{equation}
F(t)=\left\vert \left\langle \psi (N_{B},0)\right\vert e^{-iHt}\left\vert
\psi (N_{A},0)\right\rangle \right\vert .
\end{equation}%
In Fig. 5 the evolution of the state $\left\vert \psi (0)\right\rangle $ is
illustrated schematically. From the investigation of Ref. \cite{Shi}, we
know that for small $N_{A}=-N_{B}=-x_{0}$, where $N_{B}$ is the mirror
counterpart of $N_{A}$, but in large $\Delta $ limit, if we take $%
B_{0}=8\left( \ln 2/\Delta ^{2}\right) ^{2}$, $F(t)$ has the form
\begin{equation}
F(t)=\exp [-\frac{1}{2}\alpha ^{2}N_{A}^{2}(1+cos\frac{2t}{\alpha ^{2}})]
\end{equation}%
which is a periodic function of $t$ with the period $T=\alpha ^{2}\pi $ and
has maximum of 1. This is in agreement with our above analysis. However, in
quantum communication, what we concern is the behavior of $F(t)$ in the case
of the transfer distance $L\gg \Delta $, where $L=2\left\vert
N_{A}\right\vert =2\left\vert N_{B}\right\vert $. For this purpose the
numerical method is performed for the case $L=500,\Delta =2,4,6$ and $%
B_{0}=8\left( \ln 2/\Delta ^{2}\right) ^{2}\lambda $. The factor $\lambda $
determines the maximum fidelity and then the optimal field distribution can
be obtained numerically. In the Ref. \cite{songz2}, Fig. 2(a), (b) and (c)
the functions $F(t)$ are plotted for different values of $\lambda $. It
shows that for the given wavepackets with $\Delta =2,4$ and $6$, there
exists a range of $\lambda $, during which the fidelities $F(t)$ are up to $%
0.748,0.958$ and $0.992$ respectively . For finite distance, the maximum
fidelity decreases as the width of Gaussian wavepacket increases. On the
other hand, the strength of the external field also determines the value of
the optimal fidelity for a given wavepacket. There exists an optimal
external field to obtain maximal fidelity, meanwhile the period of $F(t)$
close to $T=\alpha ^{2}\pi $. This shows a difference from the ideal system,
i.e. continuous harmonic systems, in which the fidelity is independent of
the strength of the external field. Numerical results indicate that it is
possible to realize near-perfect quantum state transfer over a longer
distance in a practical ferromagnetic spin chain system.

In summary, we have shown that a perfect quantum transmission can be
realized through a universal quantum channel provided by a quantum spin
system with spectrum structure, in which each eigenenergy is commensurate
and matches with the corresponding parity. According to this SPMC for the a
mirror inversion symmetry \cite{MIS}, we can implement the perfect quantum
information transmission with several novel pre-engineered quantum spin
chains. For more practical purpose, we prove that an approximately
commensurate spin system can also realize near-perfect quantum state
transfer in a ferromagnetic Heisenberg chain with uniform coupling constant
in an external field. Numerical method has performed to study the fidelity
for the system in a parabolic magnetic field. The external field plays a
crucial role in the scheme. It induces a lower quasi-harmonic spectrum,
which can drive a Gaussian wavepacket from the initial position to its
mirror counterpart. The fidelity depends on the initial position (or
distance $L$), the width of the wavepacket $\Delta$ and the magnetic field
distribution $B(i)$ via the factor $\lambda $. Thus for given $L$ and $%
\Delta $, proper selection of the factor $\lambda $ can achieve the optimal
fidelity. Finally, we conclude that it is possible to implement near-perfect
Gaussian wavepacket transmission over a longer distance in many-body system.

\section{V. Quantum storage based on the spin chain}

Recently a universal quantum storage protocol \cite{lukin1,zoller,exp} was
presented to reversibly map the electronic spin state onto the collective
spin state of the surrounding nuclei ensemble in a quantum well (see the
Fig. 6). Because of the long decoherence time of the nuclear spins, the
information stored in them can be robustly preserved.
\begin{figure}[tbp]
\includegraphics[bb=80 350 500 600, width=7 cm, clip]{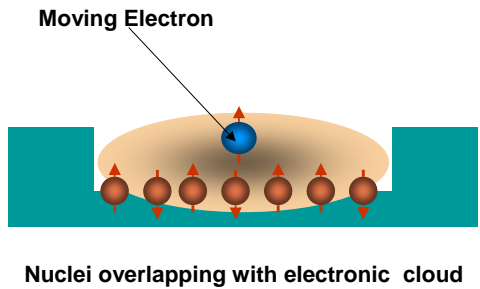}
\caption{the electronic spin state onto the collective spin state of the
surrounding nuclei ensemble in a quantum well }
\end{figure}
When all nuclei (with spin operators $I_{x}^{(i)},I_{y}^{(i)},I_{z}^{(i)}$)
of spin $I_{0}$ are coupled with a single electron spin with strength $g_{i}$%
, a pair of collective operators\cite{szs}
\begin{equation}
B=\frac{\sum_{i=1}^{N}g_{i}I_{-}^{(i)}}{\sqrt{2I_{0}\sum g_{j}^{2}}}
\end{equation}%
and its conjugate $B^{+}$ are introduced to depict the collective
excitations in ensemble of nuclei with spin $I_{0}$ from its polarized
initial state$\left\vert G\right\rangle =\left\vert -NI_{0}\right\rangle
=\prod\limits_{i=1}^{N}\left\vert -I_{0}\right\rangle _{i}$which denotes the
saturated ferromagnetic state of nuclei ensemble. There is an intuitive
argument that if the $g_{i}$s have different values, while the distribution
is "quasi-homogeneous", $B$ and $B^{\dagger }$ can also be considered as
boson operators satisfying $[B,B^{+}]\rightarrow 1$ approximately.

Song, Zhang and Sun analyzed the universal applicability of this protocol in
practice \cite{szs}. It was found that only under two homogeneous conditions
with low excitations, the many-nuclei system approximately behaves as a
single mode boson and its excitation that can serve as an efficient quantum
memory. The low excitation condition requires a ground state with all spins
orientated, which can be prepared by applying a magnetic field polarizing
all spins along a single direction. With the consideration of spontaneous
symmetry breaking for all spins orientated, a protocol of quantum storage
element was proposed to use a ferromagnetic quantum spin system, instead of
the free nuclear ensemble, to serve as a robust quantum memory.

\begin{figure}[h]
\includegraphics[width=4cm,height=4cm]{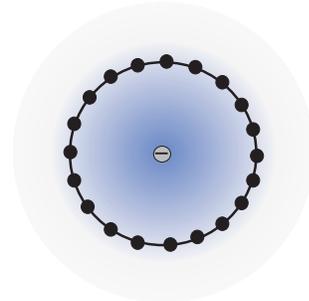}
\caption{The configuration geometry of the nuclei-electron system. The
nuclei are arranged in a circle within a quantum dot to form a ring array.
To turn on the interaction one can push a single electron towards the center
of the circle along the axis that perpendicular to the plane.}
\end{figure}

The configuration of the quantum storage element is illustrated in Fig. 7.
The nuclei are arranged in a circle within a quantum dot to form a spin ring
array. A single electron is just localized in the center of the ring array
surrounded by the nuclei. The interaction of the nuclear spins is assumed to
exist only between the nearest neighbors while the external magnetic field $%
B_{0}$ threads through the spin array. Then the electron-nuclei system can
be modelled by a Hamiltonian $H=H_{e}+H_{n}+H_{en}$. It contains the
electronic spin Hamiltonian $H_{e}=g_{e}\mu _{B}B_{0}\sigma ^{z},$the
nuclear spin Hamiltonian
\begin{equation}
H_{n}=g_{n}\mu _{n}B_{0}\sum_{l=1}^{N}S_{l}^{z}-J\sum_{l=1}^{N}\mathbf{S}_{l}%
\mathbf{\cdot S}_{l+1}  \label{**}
\end{equation}%
with the Zeeman split and the ferromagnetic interaction $J>0$, and the
interaction between the nuclear spins and the electronic spin
\begin{equation}
H_{en}=\frac{\lambda }{2N}\sigma ^{+}\sum_{l=1}^{N}S_{l}^{-}+h.c.  \label{*}
\end{equation}%
Here, $g_{e}$ ($g_{n}$) is the Lande $g$ factor of electron (nuclei), and $%
\mu _{B}$ ($\mu _{N}$) the Bohr magneton (nuclear magneton). The Pauli
matrices $S_{l}^{-}$ and $\sigma ^{+}$ represent the nuclear spin of the $l$%
-th site and the electronic spin respectively. The denominator $N$ in Eq. (%
\ref{*}) originates from the envelope normalization of the localized
electron wave-function \cite{lukin1,zoller,exp}. The hyperfine interactions
between nuclei and electron are proportional to the envelope function of
localized electron. The electronic wave function is supposed to be
cylindrical symmetric, e.g., the $s$-wave component. Thus the coupling
coefficient $\lambda \propto |\psi \left( \mathbf{r}\right) |^{2}$ is
homogenous for all the $N$ nuclei in the ring array.

To consider the low spin wave excitations, the discrete Fourier
transformation defines the bosonic operators
\begin{equation}
b_{k}=\frac{1}{\sqrt{N}}\sum_{l=1}^{N}e^{i\frac{2\pi kl}{N}}S^{-}_{l},
\end{equation}%
in the large $N$ limit. Then one can approximately diagonalize the
Hamiltonian (\ref{**}) as
\begin{equation*}
H_{T}=H_{N}+\sum_{k=1}^{N-1}\omega _{k}b_{k}^{\dag }b_{k}
\end{equation*}%
where $H_{N}$ is a Jaynes-Cummings (JC) type Hamiltonian
\begin{equation}
H_{N}=\omega _{N}b_{N}^{\dag }b_{N}+\frac{\Omega }{2}\sigma ^{z}+\lambda
\sqrt{\frac{s}{2N}}\left( \sigma ^{+}b_{N}+\sigma ^{-}b_{N}^{\dag }\right)
\end{equation}%
Then we obtain the dispersion relation for magnon or the spin wave
excitation
\begin{equation}
\omega _{k}=g_{n}\mu _{n}B_{0}+2Js-2Js\cos \frac{2\pi k}{N}.
\end{equation}%
The above results show that $H_{T}$ only contains the interaction of the $N$%
-th magnon with the electronic spin and the other $N-1$ magnons decouples
with it. Here, the frequency of the boson $\omega _{N}=g_{n}\mu _{n}B_{0}$
and the two level spacing $\Omega =2g^{\ast }\mu _{B}B_{0}$ can be modulated
by the external field $B_{0}$ simultaneously.

The process of quantum information storage can be implemented in the
invariant subspace of the electronic spin and the $N$-th magnon. Now we can
describe the quantum storage protocol based on the above spin-boson model.
Suppose the initial state of the total system is prepared so that there is
no excitation in the $N$ nuclei at all while the electron is in an arbitrary
state $\rho _{e}\left( 0\right) =\sum_{n,m=\pm }\rho _{nm}\left\vert
n\right\rangle \left\langle m\right\vert $ \ where $\left\vert
+\right\rangle $ ($\left\vert -\right\rangle $) denotes the electronic spin
up\ (down) state. The initial state of the total system can then be written
as
\begin{equation}
\rho \left( 0\right) =\rho _{b}\left( 0\right) \otimes \left\vert
0_{N}\right\rangle \left\langle 0_{N}\right\vert \otimes \rho _{e}\left(
0\right)
\end{equation}%
in terms of $\rho _{b}\left( 0\right) =\left\vert \{0\}\right\rangle
_{N-1}\left\langle \{0\}\right\vert $ where $\left\vert n_{1},n_{2},\cdots
,n_{N-1}\right\rangle \equiv \left\vert \left\{ n_{k}\right\} \right\rangle
_{N-1}$ ( $k=1$, $2$, $\cdots $, $N-1$) denotes the Fock state of the other $%
N-1$ magnons. If we set $B_{0}=0$, at $t=T\equiv (\pi /\lambda )\sqrt{N/2s}$%
, the time evolution from $\rho \left( 0\right) $ is just described as a
factorized state
\begin{equation}
\rho \left( T\right) =\rho _{b}\left( 0\right) \otimes w_{F}\otimes
\left\vert -\right\rangle \left\langle -\right\vert ,
\end{equation}%
where $w_{F}=\sum_{n,m=0,1}w_{nm}\left\vert n_{N}\right\rangle \left\langle
m_{N}\right\vert $ \ is the storing state of the $N$-th magnon with
\begin{equation}
w_{nm}=\rho _{nm}\exp \left[ \frac{i}{2}\left( m-n\right) \pi \right] .
\end{equation}%
Here, to simplify our expression, we have denoted $\rho _{++}\equiv \rho
_{00}$, $\rho _{+-}\equiv \rho _{01}$, $\rho _{-+}\equiv \rho _{10}$, $\rho
_{--}\equiv \rho _{11}$. The difference between $w_{F}$ and $\rho _{e}\left(
0\right) $ is only an unitary transformation independent of the stored
initial state $\rho _{e}\left( 0\right) $.

So far we have discussed the ideal case with homogeneous coupling between
the electron and the nuclei, that is, the coupling coefficients are the same
constant $\lambda $ for all the nuclear spins. However, the inhomogeneous
effect of coupling coefficients has to be taken into account if what we
concern is beyond the s-wave component, in which the wave function is not
strictly cylindrical symmetric. In this case, the quantum decoherence
induced by the so-called quantum leakage has been extensively investigated
for the atomic ensemble based quantum memory \cite{sun-you}. We now discuss
the similar problems for the magnon based quantum memory.

For general case, $\lambda _{l}\propto |\psi \left( \mathbf{r}_{l}\right)
|^{2}$ vary with the positions of the nuclear spins where $\psi \left(
\mathbf{r}_{l}\right) $ is the envelope function of the electron at site $%
\mathbf{r}_{l}$. In this case, the Hamiltonian contains terms other than the
interaction between the spin and $N$-th mode boson, that is, the
inhomogeneity induced interaction
\begin{equation}
V=\lambda \sqrt{\frac{s}{2N}}(\sigma _{+}\sum_{k=1}^{N-1}\chi _{k}b_{k}+h.c.)
\end{equation}%
should be added in our model Hamiltonian $H_{T}$ where $\chi
_{k}=\sum_{l=1}^{N}\frac{\lambda _{l}}{\lambda N}\exp [i2\pi kl/N]$.

For a Gaussian distribution of $\lambda _{l}$, e.g. $\lambda _{l}=(\lambda /%
\sqrt{2\pi }\sigma )\exp (-(l-1)^{2}/(2\sigma ^{2}))$with width $\sigma $
and $\lambda _{1}=\lambda $, the corresponding inhomogeneous coupling is
depicted by
\begin{equation}
\chi _{k}=\frac{1}{N}\sum_{l-1=0}^{N-1}\frac{1}{\sqrt{2\pi }\sigma }e^{\frac{%
-(l-1)^{2}}{2\sigma ^{2}}+i\frac{2\pi kl}{N}}
\end{equation}%
Fig. 8 shows the magnitude of $\chi _{k}$ for different Gaussian
distributions of $\lambda _{l}$ with different widths $\sigma $. It
indicates that the modes near $1$ and $N-1$ have a stronger coupling with
the electron. When the interaction gets more homogenous (with larger $\sigma
$) the coupling coefficients $\chi _{k}$ for all the modes from $1$ to $N-1$
become smaller. When the distribution is completely homogeneous, all the
couplings with the $N-1$ magnon modes vanish and then we obtain the
Hamiltonian $H_{T}$.

In the following we will adopt a rather direct method to analyze the
decoherence problem of our protocol resulting from dissipation. If $N$ is so
large that the spectrum of the quantum memory is quasi-continuous, this
model is similar to the "standard model" of quantum dissipation for the
vacuum induced spontaneous emission \cite{Louisell}. The $N-1$ magnons will
induce the quantum dissipation of the electronic spin with a decay rate%
\begin{equation}
\gamma =2\pi \sum_{k=1}^{N}\frac{\lambda ^{2}s|\chi _{k}|^{2}}{2N}\delta
\left( \omega _{k}-2\lambda \sqrt{\frac{s}{2N}}\right) .
\end{equation}%
Let $\left\vert \Psi \right\rangle $ be the ideal evolution governed by the
expected Hamiltonian $H_{T}$ without dissipation while the realistic
evolution $\left\vert \Psi ^{\prime }\right\rangle $ governed by the
Hamiltonian with dissipation. Suppose the initial state of the electron is $%
(\left\vert +\right\rangle +\left\vert -\right\rangle )/\sqrt{2}$, we can
analytically calculate the fidelity
\begin{eqnarray}
&&F(t)=|\langle \Psi \left\vert \Psi ^{\prime }\right\rangle |=\frac{1}{2}%
(1+e^{-\frac{\gamma }{2}t})\times   \notag \\
&&\sec \varphi (\cos gt\cos \left( \Delta _{1}^{\prime }t+\varphi \right)
+\sin gt\sin \Delta _{1}^{\prime }t),
\end{eqnarray}%
where $\varphi =\arcsin \sqrt{2N\gamma ^{2}/\lambda ^{2}s},g=\lambda \sqrt{%
s/2N}$ and $\Delta _{1}^{\prime }=\sqrt{g^{2}-\gamma ^{2}}$.

Fig. 8  shows the curve of the fidelity $F(t)$ changing with time $t$. We
can see that the fidelity exhibits a exponential decay behavior with a
sinusoidal oscillation. At the instance when we have just implemented the
quantum storage process, the fidelity is about $1-\pi \gamma /8$. Therefore,
the deviation from the ideal case with homogeneous couplings is very small
for $\gamma /g<<1$. Since the ring-shape spin array with inhomogeneous
coupling is just equivalent to an arbitrary Heisenberg spin chain in the
large $N$ limit, the above arguments means that an arbitrary Heisenberg
chain can be used for quantum storage following the same strategy addressed
above if $\gamma /g$ is small, i.e., the inhomogeneous effect is not very
strong.
\begin{figure}[h]
\includegraphics[width=5cm,height=3cm]{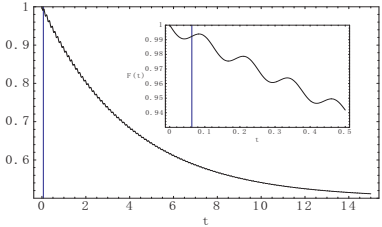}
\caption{The fidelity $F\left( t\right) $ in the large $N$ limit. The
vertical line indicates the instant\ $\frac{\protect\pi }{2g}$, at which the
quantum storage is just implemented. Here $\frac{\protect\gamma }{g}=\frac{1%
}{50}$. The inset shows the decaying oscillation with details of $F\left(
t\right) $ in a small region near the instant $\frac{\protect\pi }{2g}$. }
\end{figure}
On the other hand, if $N$ is small, the spectrum of the quantum memory is
discrete enough to guarantee the adiabatic elimination of the $N-1$ magnon
modes, i.e., $\lambda \sqrt{s/2N}\chi _{k}/|\omega _{k}|<<1$ for the $N-1$
magnon modes. As a consequence of this adiabatic elimination, the quantum
decoherence or de-phasing can result from the mixing of different magnon
modes.

\textbf{Acknowledgement:} \textit{We acknowledge\ the \
collaborations \ with P. Zhang, Yong  Li, Y.D Wang, B.Chen, \
X.F.Qian, T.Shi, Ying Li and R. Xin, which resulting in our
systematical researches on the quantum spin based quantum
information processing. SZ's work is supported by the Innovation
Foundation of Nankai university. CPS acknowledge the support of
the CNSF (grant No. 90203018), the Knowledge Innovation Program
(KIP) of the Chinese Academy of Sciences, the National Fundamental
Research Program of China (No. 001GB309310).}


\begin{thebibliography}{a}
\bibitem[a]{email} Electronic addresses: songtc@nankai.edu.cn\newline
suncp@itp.ac.cn

\bibitem[b]{www} Internet www site: http://www.itp.ac.cn/\symbol{126}suncp

\bibitem{q-inf} D. P. DiVincenzo and C. Bennet, Nature \textbf{404}, 247
(2000) and references therein

\bibitem{lukin} M. D. Lukin, Rev. Mod. Phys. \textbf{75}, 457 (2003).

\bibitem{Flei} M. Fleischhauer and M. D. Lukin, Phys. Rev. Lett, \textbf{84}%
, 5094 (2000); Phys. Rev. A, \textbf{65}, 022314 (2002)

\bibitem{sun-prl} C. P. Sun, Y. Li, and X. F. Liu, Phys. Rev. Lett. \textbf{%
91}, 147903 (2003).

\bibitem{za-store} E. Pazy, I. D'Amico, P. Zanardi, and F. Rossi, Phys. Rev.
B \textbf{64}, 195320 (2001)

\bibitem{lukin1} J. M. Taylor, C. M. Marcus and M. D. Lukin, Phys. Rev.
Lett, \textbf{90}, 206803 (2003).

\bibitem{zoller} A. Imamoglu, E. Knill, L. Tian and P. Zoller, Phys. Rev.
Lett. \textbf{91}, 017402 (2003)

\bibitem{exp} M. Poggio et al., Phys. Rev. Lett. \textbf{91}, 207602 (2003)

\bibitem{szs} Z. Song, P. Zhang and C. P. Sun, quant-ph/0409120,Effective
boson-spin model for nuclei ensemble based universal quantum memory,
submitted to Phys. Rev. B.

\bibitem{wlss} Y. D. Wang, Y. Li, Z. Song and C. P. Sun, cond-mat/0409120
Magnon based quantum storage of electronic spin with a ring array of nuclear
spins, submitted to Phys. Rev. A (2004).

\bibitem{Div} D. P. DiVincenzo, D. Bacon, J. Kempe, G. Burkard, and K. B.
Whaley, Nature 408, 339 (2000).

\bibitem{Bose} S. Bose, Phys. Rev. Lett. \textbf{91}, 207901 (2003).

\bibitem{Dagotto} E. Dagotto and T. M. Rice, Science \textbf{271}, 618
(1996).

\bibitem{White} S. White, R. Noack, and D. Scalapino, Phys. Rev. Lett.
\textbf{73}, 886 (1994); R. Noack, S. White, and D. Scalapino, Phys. Rev.
Lett. \textbf{73}, 882 (19940).

\bibitem{Matt1} M. Christandl, N. Datta, and J. Landahl, Phys. Rev. Lett.
\textbf{92}, 187902 (2004).

\bibitem{Matt2} M. Christandl, N Datta, T. C. Dorlas, A. Ekert, A. Kay and
A. J. Landahl quant-ph/0411020.

\bibitem{songz1} T. Shi, Ying Li, Z. Song, and C. P. Sun, cond-mat/0408152,
Quantum state transfer via the ferromagnetic chain in a spatially modulated
field, submitted to Phys. Rev. A (2004).

\bibitem{songz2} Ying Li, T.Shi, B. Chen, Z. Song, C.P.Sun,
quant-ph/0406159, Quantum state transmission via a spin ladder as a robust
data bus, Phys. Rev. A (2004) in press.

\bibitem{geo} C. P. Sun, P. Zhang and Y. Li,quant-ph/0311052, Geometric
Quantum Information Storage Based on Atomic Ensemble ; Y. Li, P. Zhang, P.
Zanardi, and C. P. Sun Phys. Rev. A 70, 032330 (2004)

\bibitem{MIS} Claudio Albanese, Matthias Christandl, Nilanjana Datta, Artur
Ekert, quant-ph/0405029.

\bibitem{QD array} D. Loss and D. P. DiVincenzo, Phys. Rev. A \textbf{57},
120 (1998); B. E. Kane, Nature (London) \textbf{393}, 133 (1998).

\bibitem{Lieb} E. Lieb, Phys. Rev. Lett. \textbf{62}, 1201 (1989); E. Lieb
and D. Mattis, J. Math. Phys. \textbf{3}, 749 (1962).

\bibitem{Song} Z. Song, Phys. Lett. A \textbf{231}, 135 (1997); Z. Song,
Phys. Lett. A \textbf{233}, 135 (1997). (2004)

\bibitem{Shn} Yu. Makhlin, G. Schon, and A. Shnirman, Rev. Mod. Phys. 73,
357 (2001)

\bibitem{Shi} T. Shi, B. Chen, Z. Song and C.P. Sun, On the harmonic
approximation for large Josephson junction coupling charge qubits, Comm.
Theor. Phys. (2004),in press.

\bibitem{sun-you} C. P. Sun, S. Yi, L. You, Phys. Rev. A \textbf{67}, 063815.

\bibitem{Louisell} W.H. Louisell, "Quantum Statistical Properties of
Radiation", John Wiley and Son's, New York, (1990).
\end{thebibliography}
\end{document}